\documentclass[11pt]{article} 
\usepackage{hyperref} 
\begin{document} 
\title{Locomotion of {\textit {C. elegans}} in structured environments}
\author{Trushant Majmudar, Eric Keaveny, Mike Shelley, Jun Zhang\\ \\\vspace{6pt} Courant Institute of Mathematical Sciences\\ New York University, New York, NY 10012}
\maketitle
\begin{abstract} 

Undulatory locomotion of microorganisms like soil-dwelling worms and spermatozoa, in structured environments, is ubiquitous in nature. They navigate complex environments consisting of fluids and obstacles, negotiating hydrodynamic effects and geometrical constraints. Here, we show fluid dynamics videos of experiments and simulations of {\textit {C. elegans}} moving in an array of micro-pillars. In addition, we show a video of transition from swimming to crawling in drop of {\textit {C. elegans}}, where the fluid is wicking into agar.
\end{abstract}

\section{Introduction} 

Low Reynolds number locomotion of undulating microorganisms, particularly in fluids with embedded microstructure, is studied here with the soil-dwelling nematode {\textit {C. elegans}} as a model organism. We show experimental data for {\textit {C elegans}} swimming in arrays of PDMS micro-pillars in square lattice, which is filled with aqueous buffer. We also show data from simulations of a self-locomoting chain of beads in a lattice of circular obstacles. The simulations include hydrodynamics, contact interactions and geometrical constraints, with an imposed torque wave on the bead-chain. The model is purely mechanical, with no sensing and behavior included. 
The nematodes are about $1$ mm in length, and $60~\mu$m thick. The lattice consists of pillars, which are $300~\mu$m wide and $300 \mu$m tall. The center-to-center distance between the pillars varies from $380~\mu$m to $610 \mu$m.
The typical swimming speed of a worm in an aqueous buffer is $U \sim 0.1$ mm/s, which implies that the Reynolds number $Re = U L / \nu \sim 0.1$, where $\nu$ is the kinematic viscosity of the fluid. We find that the nematodes exhibit a rich array of dynamical behaviors while swimming in structured environments. We also find that a purely mechanistic model is able to reproduce these behaviors successfully. 
Following is an explanation of the set of videos we show here.

\begin{enumerate}
\item The first video shows the locomotion of {\textit {C. elegans}} (experiment) and simulations of a bead-chain in a lattice, where they are in constant contact with the pillars. Both videos show enhanced locomotion as they move diagonally, by pushing against the pillars. The observed velocity in the lattice is about an order of magnitude higher than the free swimming velocity ($V_{diag} = 0.8$ mm/s). Moreover, the simulations also capture the gait of the real nematode.

\item The second video shows an experiment and the corresponding simulation, where both the nematode and the bead-chain exhibit closed circular trajectories. In the experiments, the nematode exits one circular trajectory to form another. In the simulations, the bead-chain is locked indefinitely on the closed periodic trajectory.

\item The last video shows a type of collective effect in structureless fluid; a large number of nematodes are in a drop of buffer placed on an agar plate. The fluid slowly seeps into agar, reducing the volume available for the nematodes to swim. As the fluid volume decreases, they collect together and exhibit a transition from swimming to crawling. 

\end{enumerate}

\end{document}